\documentclass[%
 reprint,
 amsmath,amssymb,
 aip,jcp,
]{revtex4-2}

\usepackage{graphicx}
\usepackage{dcolumn}
\usepackage{bm}

\usepackage{subfigure}

\usepackage{graphicx}%
\usepackage{dcolumn}%
\usepackage{bm}%
\usepackage{hyperref}%

\usepackage{amssymb}
\usepackage{tikz}

\usetikzlibrary{shapes.callouts}
\usetikzlibrary{shapes.geometric,shapes.arrows,decorations.pathmorphing}
\usetikzlibrary{snakes,matrix,chains,scopes,positioning,fit,backgrounds}

\usepackage{wasysym}
\usepackage[version=4]{mhchem}

\usetikzlibrary{shapes.callouts}
\usetikzlibrary{shapes.geometric,shapes.arrows,decorations.pathmorphing}
\usetikzlibrary{matrix,chains,scopes,positioning,arrows,fit}

\linethickness{0.4mm}
\definecolor{mma1}{rgb}{0.3725,0.5098,0.7020}
\definecolor{mma2}{rgb}{0.8745,0.6078,0.2039}
\definecolor{mma3}{rgb}{0.507813,0.714844,0.2039}
\definecolor{mma4}{rgb}{0.9137,0.3882,0.2398}

\tikzset{
  laser1/.style   = { ultra thick, mma4,decorate, decoration={snake}},
  laserI/.style = {ultra thick, mma1,decorate, decoration={snake}},
  laserIp/.style = {ultra thick, mma1, decorate, decoration={snake}},
  laserS/.style = {ultra thick, mma2,decorate, decoration={snake}},
  laserSp/.style = {ultra thick, mma2, decorate, decoration={snake}},
  laser2/.style   = { ultra thick,blue},
  connect/.style = { dashed, red },
  notice/.style  = { draw, rectangle callout, callout relative pointer={#1} },
  label/.style   = { text width=2cm }
}

\begin{document}
\title{Concerning the stability of
biexcitons in hybrid HJ aggregates of $\pi$-conjugated polymers}
\author{Eric~R.~Bittner}
\email{ebittner@central.uh.edu}
\affiliation{Department of Chemistry, University of Houston, Houston, Texas 77204, United~States}

\author{Carlos~Silva}
\affiliation{School of Chemistry and Biochemistry, Georgia Institute of Technology, 901 Atlantic Drive, Atlanta, GA~30332, United~States}
\affiliation{School of Physics, Georgia Institute of Technology, 837 State Street, Atlanta, GA~30332, United~States}
\affiliation{School of Materials Science and Engineering, Georgia Institute of Technology, North Avenue, Atlanta, GA~30332, United~States}

\date{\today}

\begin{abstract}
 Frenkel excitons are the primary photoexcitations in organic semiconductors and are
 ultimately responsible for the optical properties of such materials. 
They are also  predicted to form \emph{bound} exciton pairs, termed biexcitons,
which   
 are consequential intermediates in a wide range of photophysical processes.
Generally, we think of bound states as arising 
from an attractive interaction.  However,
here we report on our recent theoretical analysis
predicting the formation of stable biexciton states
in a conjugated polymer material arising from 
both attractive and repulsive interactions.
We show that in J-aggregate systems, 2J-biexcitons
can arise from repulsive dipolar interactions with energies $E_{2J}> 2E_J$ while in H-aggregates, 2H-biexciton states $E_{2H} < 2E_H$ corresponding to attractive dipole exciton/exciton interactions.  
These predictions are corroborated by using ultrafast double-quantum coherence spectroscopy on a PBTTT material that exhibits both J- and H-like  excitonic behavior.
\end{abstract}

\maketitle

\section{Introduction}

It is generally understood that the
primary photoexcitations in organic semiconducting 
materials 
are molecular $\pi-\pi^*$ electronic singlet states (\ce{S_1}) termed Frenkel excitons~\cite{frenkel1931transformation}. 
While local in nature, at
sufficiently high packing densities, 
excitons can delocalized over several molecular units
and sufficiently higher excitation densities, 
exciton-exciton interactions begin to dominate the optical properties of such materials~\cite{agranovich1968collective}. 
Biexcitons, bound pairs of excitons, 
 are consequential intermediates in a wide range of photophysical processes such as exciton dissociation into electrons (e$^{-}$) and holes (h$^+$) ~\cite{silva2001efficient},
 \begin{align}
\ce{S0 + 2 $\hbar\omega$ -> [2 S1]$^\ddagger$ -> 2e^- + 2h^+}
\end{align}
bimolecular annihilation~\cite{stevens2001exciton},
\begin{align}
\ce{S1 + S1 -> [2 S1]$^\ddagger$ -> S0 + S0}
\end{align}
and singlet fission producing triplet (\ce{T1}) states ~\cite{silva2002exciton}
\begin{align}
\ce{S0 + 2 $\hbar\omega$ -> [2 S1]$^\ddagger$ -> T1 + T1}. 
\end{align}
Ref.~\citenum{stevens2001exciton} notes that bimolecular annihilation may be mediated both by resonance energy transfer and diffusion-limited exciton-exciton scattering, but in either case we invoke the key 
intermediate \ce{[2 S1]$^\ddagger$}.
Examples of this occur 
 in biological light harvesting complexes where multi-exciton interactions may play important roles~\cite{scholes2011lessons}
 in the excitonic transport process,and biexcitons can be crucial in cascade quantum emitters as a source of entangled photons~\cite{n2021mechanistic}. 
While ample theoretical work points towards the existence of biexcitons in organic solids~\cite{spano1991biexciton,Vektaris1994,guo1995stable,gallagher1996theory,mazumdar1996exciton,agranovich2000kinematic,kun2009effect}, 
and in optical lattices~\cite{xiang2012tunable}, there has been only indirect evidence of the dynamic formation of two-quantum exciton bound states in polymeric semiconductors by incoherent, sequential ultrafast excitation~\cite{silva2001efficient,stevens2001exciton,silva2002exciton,chakrabarti1998evidence,klimov1998biexcitons}. 

\begin{figure}
    \centering
\begin{tikzpicture}
\begin{scope}
\node at (-4,1.75) {\sf\Large (a)};
          \node at (0,0) {\includegraphics[]{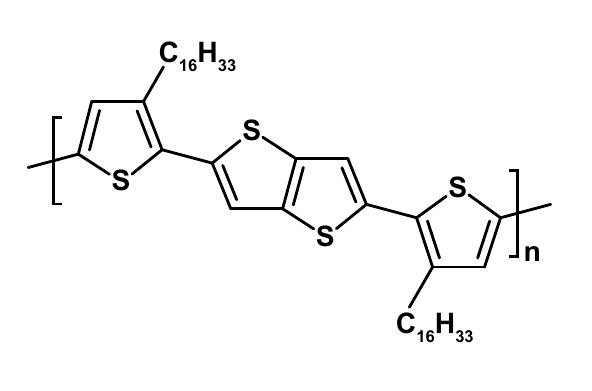}};
\end{scope}

\begin{scope}[scale=1.,shift={(-3,-3)}]
\node at (-1,1.25) {\sf \Large (b)};

\foreach \x [count=\n]in {0,1,...,5}{ 
    \filldraw[rotate=0,mma1!90,rotate around={30:(\x,0)}] (\x,0) ellipse (5mm and 2.5mm);
    \draw[->,thick,yellow,rotate around={30:(\x,0)}](-0.25+\x,0)--(0.25+\x,0);
    
    \filldraw[rotate=0,mma1!90,rotate around={30:(\x,-1)}] (\x,-1) ellipse (5mm and 2.5mm);
    \draw[->,thick,yellow,rotate around={30:(\x,-1)}](-0.25+\x,-1)--(0.25+\x,-1);
}
    
    \end{scope}          
    \end{tikzpicture}
    
    \caption{(a) Chemical structure of the PBTTT polymer. (b) Local packing of excitonic
    units (ellipsoids) with superimposed transition dipoles (yellow arrows)
    with J-like interactions along the horizontal axis and H-like interactions along the vertical axis.}
    \label{fig:1}
\end{figure}
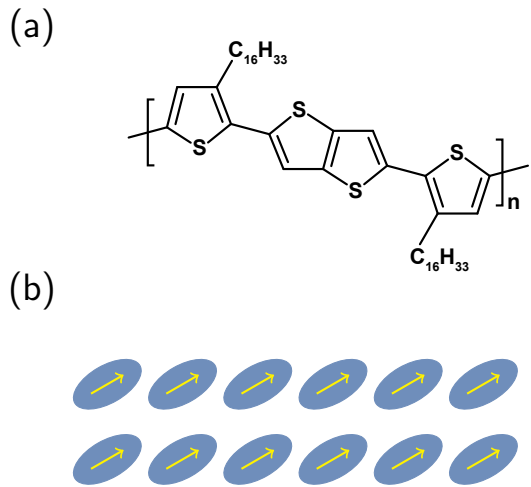

Recently, we reported upon the 
the \emph{direct} spectroscopic observation of \emph{bound} Frenkel biexcitons, {\em i.e.}, bound two-exciton quasiparticles (\ce{[2S1]$^\ddagger$}), in a  model polymeric semiconductor,  [poly(2,5-bis(3-hexadecylthiophene-2-yl)thieno[3,2-b]thiophene)]~\cite{mcculloch2006liquid} (PBTTT) 
using coherent two-dimensional ultrafast spectroscopy.
\cite{meza2021frenkel}
The chemical structure of PBTTT is given in Fig.~\ref{fig:1}(A).
PBTTT is unique in that depending upon processing conditions, it can support the 
formation of both H and J aggregate single exciton states, suggesting 
an arrangement as sketched in Fig\ref{fig:1}(B) in which intra-chain
J-like excitons can form along the chains spanning over several PBTTT subunits, 
while inter-chain H-like excitons can form due to parallel stacking of several 
chains within the aggregate. 
The experimental observations revealed a correlation 
between peaks in the single and double 
quantum spectra the correspond to the formation of
2H and 2J biexciton species.  This conclusion was supported
by both a computational model and theoretical analysis based
upon a quasi-one-dimensional continuum model.

Here present an overview of
the theoretical model we developed for biexcitons and use 
it to discuss biexcitons in 
related organic polymer materials.   First, we show how one can 
reduce the two-dimension lattice problem into two separate one-dimensional 
problems and  use a Greens 
function approach to account for the contact interaction between excitons. 
This gives the criterion for the overall stability of the 
bound biexciton states in terms of the exciton band-width
and contact interaction.  Formally, the lattice model reduces to a
one-dimensional continuum model with $\delta$-function potential.
We append to this ``text-book'' model a deformable classical media
to examine the contribution of the lattice reorganization about the 
bound-biexciton states and find that this stabilizes
the attractively bound biexciton state but destabilizes the repulsively bound
states.  We conclude by discussing the experimental observations
and the need for  a better 
theoretical understanding of bound biexciton states.

\section{Theoretical model}

\subsection{Homogeneous lattice model in 1- and 2-dimensions}
To explore the possibility of having multiple species
of bound, biexcitons in the same system, 
we begin by writing a generic lattice model for the system
by defining exciton operators $a_\textbf{n}$ and $a_{\textbf{n}}^\dagger$ 
\begin{align}
    a_\textbf{n} a_\textbf{m}^\dagger - (-1)^{\delta_{\textbf{nm}}}a_\textbf{m}^\dagger a_\textbf{n} = \delta_{\textbf{nm}}.
\end{align}
These operators are Paulion operators that
create and remove single excitations
on to a site labeled by $\textbf{n}$.  On a given site, they obey
the Fermion relation $a_\textbf{n}a_\textbf{n}^\dagger + a_n^\dagger a_n=1$
which enforces Pauli exclusion since 
$a_na_n =a_n^\dagger a_n^\dagger  = 1$, 
but commute ``off-site'' with 
$[a_\textbf{n},a_\textbf{m}^\dagger] = 0$ when $\textbf{n}\ne \textbf{m}$. This 
is different than the usual Fermion algebra
where the anti-commutation rule is applies over all 
sites, ultimately giving rise to the exchange interaction. 
Further, we can write a generic multi-exciton Hamiltonian 
as
\begin{align}
    H = \sum_{\textbf{nm}} h_{\textbf{nm}}a_\textbf{n}^\dagger a_\textbf{m}
    + \frac{1}{2}\sum_{\textbf{nm}}
    U_{\textbf{nm}}a^\dagger_\textbf{n}a^\dagger_\textbf{m} a_\textbf{n} a_\textbf{m}.
    \label{eq:hamiltonian}
    \end{align}
    where $h_{\textbf{nm}}$ describes the single-exciton 
    dynamics and $U_{\textbf{nm}}$ is the exciton/exciton interaction.
    In principle, the parameters entering into 
the Hamiltonian in Eq.~\ref{eq:hamiltonian} are defined by the system of interest. For the case of excitons, 
the diagonal elements of the single particle term defines the energy to place an exciton 
in site $\textbf{n}$, and we write $h_{\textbf nn} = \epsilon_n$.  For a homogeneous lattice, 
all site energies are the same, and $\epsilon_n = \epsilon_o$.  Similarly, the off-diagonal 
elements of $h_{\textbf{nm}}$ correspond to the 
matrix elements for transferring an excitation
from site $\textbf{n}\to \textbf{m}$.  
To a good approximation, the single-exciton 
transfer interaction can be described
within the dipole-dipole approximation 
as described above. 
    This model differs from the Hubbard model commonly studied in 
    condensed matter physics in that we explicitly exclude double occupancy of 
    each lattice and the exciton/exciton interaction is taken to be between occupied
    neighbors. 
Formally, a Frenkel exciton corresponds to a single electron/hole excitation on a given site. However,  molecules are not point particles and excitons may acquire some intramolecular charge-transfer character. 
Therefore, we anticipate that $U_{\textbf{nm}}$
is also dipole-dipole like and reflects the 
relative orientation of the {\em static} exciton dipole moments.

For a one-dimensional chain with lattice spacing $a$, $n$ is simply an index
along the chain such that the site location is given by $r_\textbf{n} = \textbf{n} a$. However, for 2- and 3-dimensional 
systems, we shall take it as an n-tuple index specifying the
site location.   
For the single particle term, $h_{\textbf{nn}}$ is the excitation 
energy for single site ($\varepsilon_\textbf{n}$) 
and $h_{\textbf{nm}}$ (for $\textbf{n}\ne \textbf{m}$)
corresponds to the hopping integral between sites. 
Upon transforming into the reciprocal space, 
\begin{align}
    \tilde a_\textbf{k} = \frac{1}{\sqrt{N}}\sum_\textbf{n} e^{-\textbf{k}\cdot \textbf{n}}a_\textbf{n}
\end{align}
one finds the single particle energy dispersion 
as
\begin{align}
    (\varepsilon(\textbf{k}) - E) e^{i\textbf{k}\cdot \textbf{n}} = \sum_\textbf{m} h_{\textbf{nm}}e^{i\textbf{k}\cdot \textbf{m}}.
\end{align}
To determine the 2-exciton states, we begin 
by writing
\begin{align}
    |\psi\rangle =\sum_{\textbf{kk}'}c_{\textbf{kk}'}\tilde a^\dagger_\textbf{k} \tilde a_{\textbf{k}'}^\dagger.
\end{align}
where $c_{\textbf{k}\textbf{k}'}$
are the expansion coefficients for this 
state.  At this point, there are various 
approaches one can take to find the 
general solutions for the Schr\"odinger
equation for the 2-exciton system. Indeed, 
for a small enough lattice, one can simply 
directly
diagonalize the Hamiltonian in Eq.~\ref{eq:hamiltonian} for a
finite sized grid.  However, we are not interested
the full solution of this problem.  Rather, we 
are focused upon only solutions corresponding to 
 \textit{bound} exciton pairs, and especially
 those bound pairs that retain their J- or H-like 
 excitonic character. 
 With this in mind, 
 we develop an analytical 
 solution that naturally 
 extends to the full 
 model.

\begin{figure}
    \centering
    \includegraphics[width=\columnwidth]{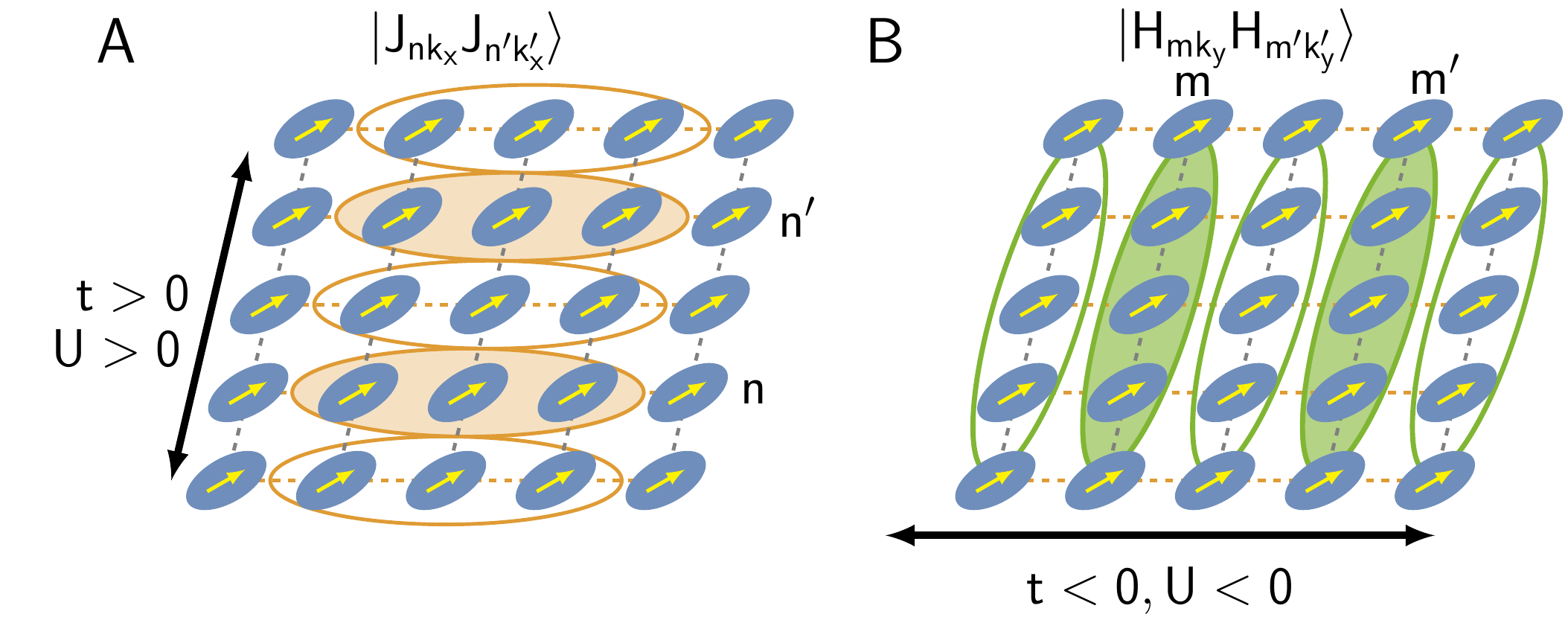}
    \caption{Sketch of 2D lattice model 
    with superimposed transition dipole for 
    each site. In (A), we define a J-aggregate
    basis and create 2J biexciton configurations
    along each row. In (B), we define an
    H-aggregate basis along each column.  As discussed in the text, these are equivalent 
    representations of the full 2D problem and are useful for reducing the 2D problem into two separate (but formally identical) 1D problems. 
    }
    \label{fig:2}
\end{figure}

\subsection{Local exciton approximation}
 In Ref. 
~\citenum{meza2021frenkel}
employed both direct diagonalization and a 
lattice Greens function approach developed by Vektaris\cite{Vektaris1994}
to study the properties of the biexciton including its dispersion 
and the affects of local disorder.  A key assumption in our model is that
we can define two equivalent quasi-one-dimensional representations for 
H-like or J-like excitons.  
For this, let us define a new set of exciton operators, 
$\hat J^\dagger_{k_x}(n)$ 
and $\hat J_{k_x}(n)$
which creates or removes an exciton with wavevector $k_x$ in the $x$-direction
localized on the $n$th row of sites. Similarly, we define operators  $\hat H_{k_y}^\dagger(m)$ and 
$\hat H_{k_y}(m)$ which create and remove excitons with wavevector $k_y$ 
in the $y$-direction, but localized to the $m$th column, 
as sketched in Fig.~\ref{fig:2}.
These can be written in terms
of the original lattice operators
\begin{align}
    \hat J_{m,k_x} = \frac{1}{\sqrt{N}}\sum_n e^{ik_xn}a_{(n,m)}\\
    \hat H_{n,k_y} = \frac{1}{\sqrt{N}}\sum_m e^{ik_{y} m}a_{(n,m)}.
\end{align}
Both are equivalent representations and we 
can choose to use either (but not both) to rewrite the
original problem in this new representation.

Thus, we can write
\begin{align}
    \sum_{\textbf{nm}}h_{\textbf{nm}}a^\dagger_{\bf n}a_{\bf m}
    &=\sum_{k_xmm'}(\varepsilon_J(k_x)\delta_{mm'} + t_H(mm')\hat J^\dagger_{m,k_x}\hat J_{m',k_x} \nonumber 
    \\
    &= \sum_{k_y nn'}(\varepsilon_H(k_y)\delta_{nn'} + t_J(nn'))\hat H^\dagger_{n,k_y}\hat H_{n'k_y} 
\end{align}
where in the first line we diagonalized in the J-direction and in the 
second, we diagonalized in the H-direction.  This implies that we can 
think of a J-exciton state as moving in the H-direction with hopping 
integral $t_H$ and H-exciton states as moving in the J-direction with 
hopping integrals $t_J$. 
The dispersion relations are then as usual
\begin{align}
    \varepsilon_J(k_x) &= \epsilon_0 + 2 t_J \cos(k_x) \\
    \varepsilon_H(k_y) &= \epsilon_0 + 2 t_H \cos(k_y) 
\end{align}
where $t_J < 0$ is the nearest-neighbor coupling along the
$x$-direction and $t_H>0$ is the nearest-neighbor coupling
taken in the $y$-direction.  
To remind, throughout we 
are taking the wavevector $k\in [-\pi,\pi]$.
Since the problem is formally separable between $x$ and $y$ 
directions, the single-particle terms do not mix 
wavevector components since $k_x$ and $k_y$ are ``good'' 
quantum numbers for this system.  
Using the $\hat J$ and $\hat H$ operators, we can reduce the 2-particle/2-dimensional problem into a 
one for pair of particles within a 1 dimensional  frame without loss of generality.

\begin{figure}
\centering
\subfigure{ \includegraphics[width=0.48\columnwidth]{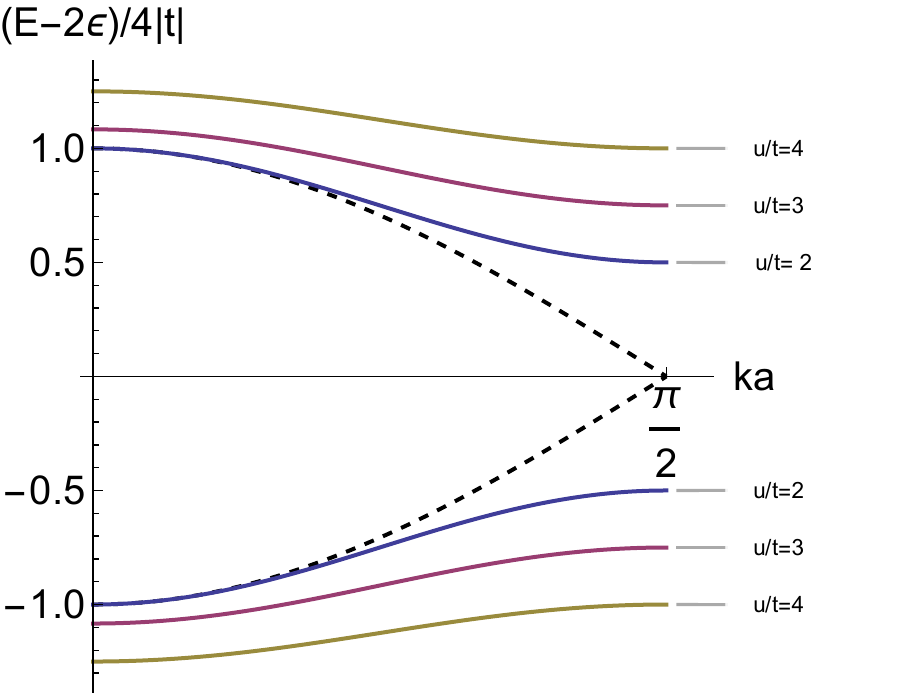}}
\subfigure{ \includegraphics[width=0.48\columnwidth]{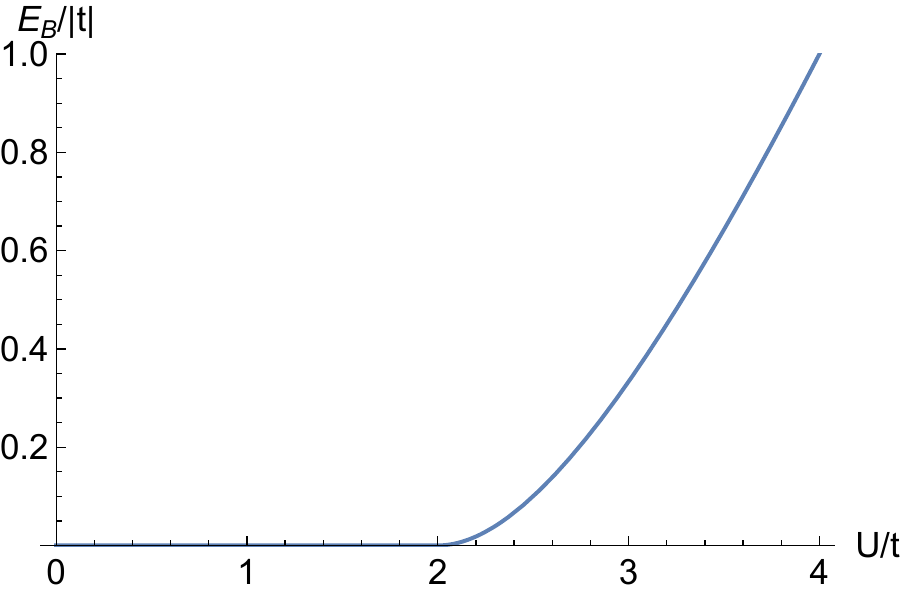}}
    \caption{
    (a) Energy bands for attractive (2H) and repulsively (2J)  bound biexcitons. Dashed lines
    correspond to the non-interacting $U\to 0$ limit of the model.  
    Solid curves are plotted for the case of $U/2t > 1$ as indicated.
    The dispersion curves need to be taken
at the corresponding points in $k$-space with a corresponding shift in the energy origins for 
each exciton species. 
(b)Variation in the biexciton binding energy at $k=0$ with increasing exciton/exciton interactions.
    }
    \label{fig:3}
\end{figure}

This decomposition suggest that 
motion in one direction can be very different than motion in 
the perpendicular direction due to the orientations
of the transition dipoles between neighboring units. 
In Fig.~\ref{fig:1}(B) we suggest how this can 
be accomplished within the context of molecular aggregates with 
$\pi$- stacking.  
Here, excitonic sites are denoted as blue ellipsoids  along 
with their respective
local transition dipole moments (yellow arrows). 
Along the $x$-direction of the 
2-dimensional lattice sketched here, the transition moments
are oriented more or less in a ``head-to-tail'' arrangement
producing a hopping matrix element in $h_{\textbf{nm}} < 0$.
In the $y$-direction, however, the transitions moments 
are aligned co-facially producing a single-particle hopping
matrix element $h_{\textbf{nm}} > 0$.  In the former case, 
the optically bright state occurs at the bottom of the 
energy band (J-aggregate), whilst in the latter case
the optically bright state occurs at the top of the energy
band (H-aggregate). The PBTTT material is unique in that
both J- and H- aggregate states can be readily observed, 
depending upon the sample preparation.

If we use the  $J_{km}$ (or $H_{kn}$) 
states as a basis for a given value of $k$, we can use 
the Greens function approach to 
find the biexciton energies as 
\begin{align}
    E_{XX}(k) = 2 \varepsilon_x + \frac{4 t^2}{U}\cos^2(k) + U
\end{align}
where $\epsilon_x$ is the excitation energy of 
either the $J$ or $H$ single exciton and $U$ is the contact 
interaction. \cite{Vektaris1994}
This expression is predicated upon $|U/2t| > 1 $ in order for the
the biexciton wavefunction to decay exponentially with exciton-exciton 
separation. 
These dispersions are plotted in 
Fig.~\ref{fig:3}a for both 2H and 2J biexcitons. In each case, the
energy origin should be shifted to twice the J or H exciton energy.
At $k=0$, the difference between the interacting and localized excitons is 
the contact energy $U$, which defines the binding energy for the exciton pair.
The binding energy must be at least greater in magnitude than $4t$, else the lowest energy
interacting state will be still within the band for the freely dissociated pairs. 
These bands will split from the freely
dissociated bands once $U/2t >1 $.

\subsection{Optical Transitions}

It is important to keep in mind that two types of  excitons sketched in Fig.\ref{fig:1} 
occur at different wave-vectors.  In Fig.\ref{fig:1}(a) the J-like exciton
has its longest wavelength in the $x$-direction (corresponding to $k_x=0$)
while having the shortest possible wavelength in the $y$-direction (i.e. $k_y = \pm \pi/a$).
Transitions to this state from the exciton vacuum (ground states) are allowed
under the dipole approximation for the transition moment.  Consequently, double excitations
can create either pairs of free J-like excitons or bound 2J excitons. 
Similarly,  optically allowed H-like single excitons are delocalized along the $y$-direction 
but localized along the $x$-direction (i.e. ${\bf k} = \{\pm\pi/a,0\}$) implying that the
double excitations must also occur at these wave-vectors. 
Bearing this in mind, the dispersion curves plotted in Fig.~\ref{fig:3} need to be taken
at the corresponding points in $k$-space with a corresponding shift in the energy origins for 
each exciton species.

\subsection{Biexciton stability}

According to our model, Frenkel biexcitons mix J-like and H-like character in terms of their collective 
quantum behavior with the requirement that the ratio of the exciton/exciton interaction and the
 perpendicular hopping term be $U/t > 0 $ which gives rise to localized biexciton states in the 
perpendicular direction.   For the 1D $\delta$-function potential, any attractive interaction with $U/t>0$ produces a localized state with localization length  $\lambda = \kappa^{-1} =2t/U$.\footnote{Note, that here
 $t = -\hbar^2/2\mu$  carries units of $[{\rm Energy}]\cdot [{\rm length}]^2$ and $U$ carries units of $[{\rm Energy}]\cdot[{\rm length}]^{-1}$ 
 so that $\lambda$ carries the appropriate units of [length].}
For the 1D lattice, bound biexciton states occur outside the band for free biexcitons. 
To gain further insight into the stability of these states, we turn 
to a continuum model  and work in a relative coordinate reference frame
where $x = |r_1 -r_2|$ is the separation between two localized excitons. 
Thus, the biexciton Schr\"odinger equation can be approximated as 
\begin{align}
    t\psi'' + U\delta(x)\psi = E\psi
\end{align}
where $U$ is the contact interaction between the two excitons.  
For bound states, $\psi(x)$ must vanish as $x\to\pm\infty$, giving that
\begin{align}
    \psi(x) = \left\{
    \begin{array}{cc}
      \sqrt{\kappa}e^{-\kappa x}   &  x > 0\\
       \sqrt{\kappa}e^{+\kappa x}  & x < 0 
    \end{array}
    \right.
    \label{eq:20}
\end{align}
where $\kappa = U/2t$ is a positive constant and $E = t \kappa^2$. In general, we take 
$t = -\hbar^2/2\mu_{eff}$ and $U<0$ for an attractive potential giving rise to a bound state
energetically {\em below} the continuum for the scattering states.  

\subsubsection{Lattice reorganization in the impurity model}

Generally speaking, one cannot discount the role of lattice reorganization 
and relaxation when discussing excitons and biexcitons in 
organic polymer semiconducting systems. 
To study this, we append to the 1D impurity model a 
continuum model for the medium a term coupling the biexciton to the lattice 
as per the Davydov model. \cite{SCOTT19921,Edler2004DirectOO,PhysRevB.82.014305,doi:10.1063/1.3592155,Goj:10,GEORGIEV2019257,GEORGIEV2019275,PhysRevB.35.3629,DAVYDOV1977379,DAVYDOV1973559}
The resulting equations of motion read
\begin{align}
i\hbar \dot \psi(x) &=  (t \nabla^2  + U\delta(x) + (E_o + 2 \chi \nabla u(x))) \psi(x) \nonumber \\ 
\ddot u - \frac{k}{m}\nabla^2 u &= 2\frac{\chi}{m}\nabla|\psi|^2
\end{align}
where $u(x)$ is the lattice deformation, $\chi$ describes the linear coupling between the biexciton and the lattice, $m$ is the 
mass of the lattice ``atoms'' and $k$ is the elastic modulus. 
If we seek traveling wave solutions, $u(x,t) = u(x-vt)$ where $v$ is the group velocity, 
we find a closure relation
\begin{align}
\nabla u = -\frac{2\chi}{k(1-(v/c)^2)} | \psi|^2
\end{align}
that gives us a non-linear Schr\"odinger equation
\begin{align}
i\hbar \dot \psi = (t\nabla^2 + g|\psi|^2  +U\delta(x))\psi
\end{align}
where $g = - 4\chi^2/(k(1-(v/c)^2))$ and $c$ is the sound velocity. 
Note that $E_o$ is a constant given by 
\begin{align}
    E_o = E - 2t +\frac{1}{2}
    \int_{-\infty}^{\infty}
    \left( m \ddot u^2 + k u'' \right)dx.
\end{align}
that we can ignore for purposes of this analysis.
The $\delta$-function potential implies the wave function 
should have the form in Eq.~\ref{eq:20}. 
Taking $\kappa$ as a variational parameter and minimizing the 
total energy, one obtains
\begin{align}
\kappa = \frac{U}{2t} + \frac{g}{8t}.
\end{align}
 $\kappa > 0$ is necessary to produce a localized state
 and from above  $U/t > 0$  and $g<0$ from its definition above, 
we have a stability requirement that if $U>0$ and $t>0$, then $-g< 4U$. 
Solving for the binding energy, 
\begin{align}
E_B = \frac{(4U + g)^2}{64 t}
\end{align}
we obtain a straight-forward estimate of the contribution of both the 
lattice and the exciton/exciton coupling to the biexciton binding. 

\begin{figure}
\includegraphics[width=\columnwidth]{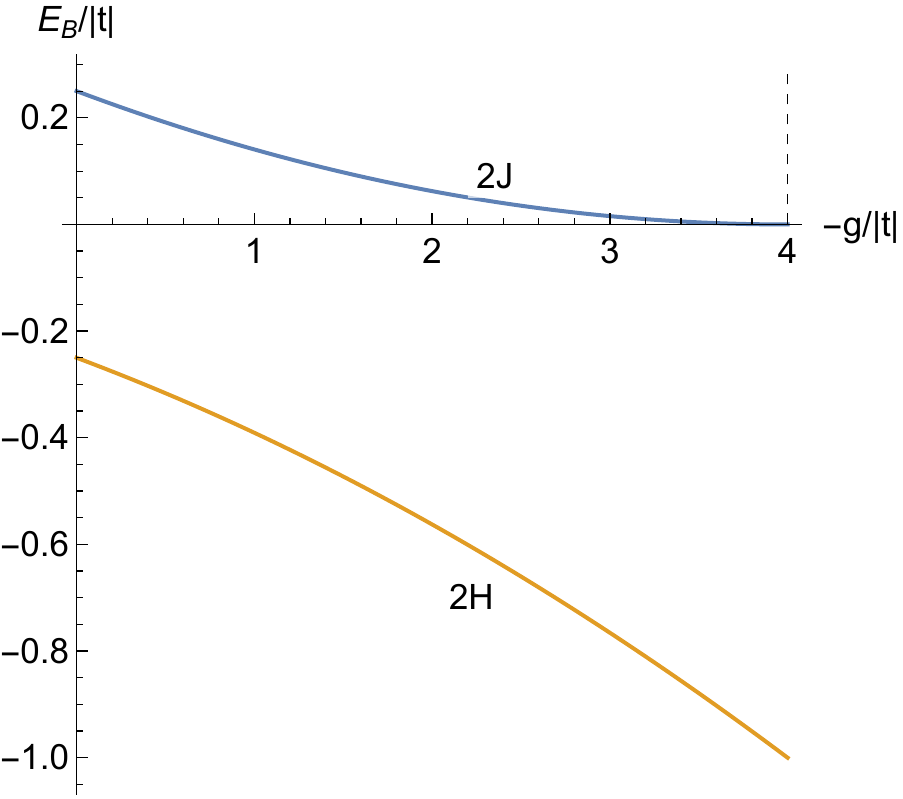}
\caption{Biexciton binding energy (in reduced units) for the 2J and 2H cases versus 
increasing exciton/lattice coupling.   For the attractively bound 2H, lattice reorganization
is expected to stabilize the biexciton state while destabilizing the 2J.
The dashed line indicates the limit of stability for the 2J case.}
\label{fig:stab}
\end{figure}

In Fig.~\ref{fig:stab} we plot the biexciton binding energy versus the non-linearity parameter, $g$.
For the attractively bound 2H, lattice reorganization
is expected to stabilize the biexciton state by further localizing the 
state ($\kappa$ increases as $g$ increases in magnitude).  On the other hand, 
for the 2J state, increasing the magnitude  of $g$ decreases $\kappa$ and 
destabilizes the otherwise bound 2J state.  When $-g = 2U$ the state is fully delocalized and
further increases in the lattice coupling lead to unbound solutions.  

\section{Comparison to experimental measurement}

\begin{figure}
    \centering
    \includegraphics[width=\columnwidth]{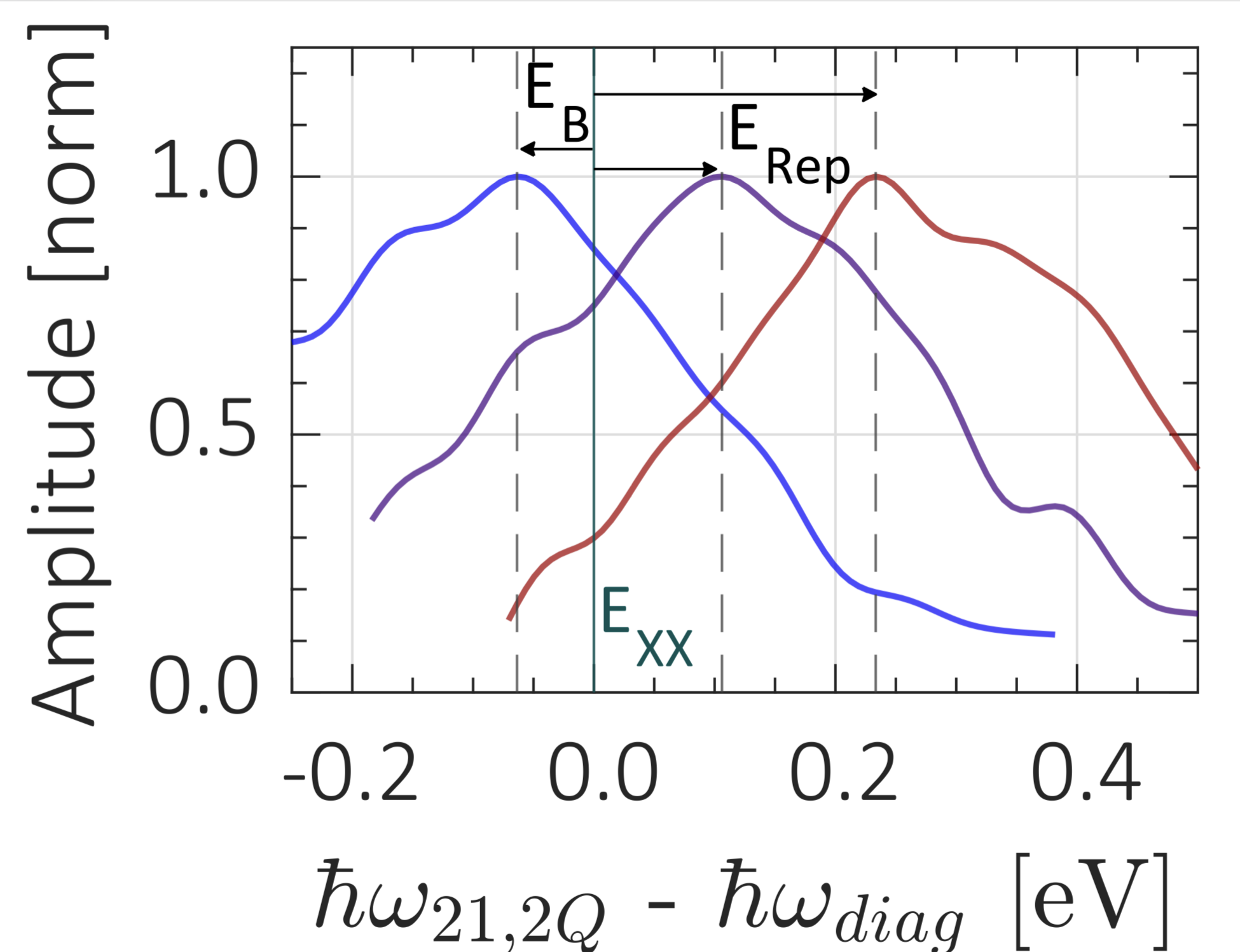}
    \caption{Spectral cuts along the two-quantum coherence energy axis, relative to the two-quantum diagonal energy axis ($ \hbar \omega_{2Q} = 2\hbar \omega_{1Q}$, $E_{XX}$) at fixed $\hbar \omega_{43} = 2.06$ (H-aggregate resonance, blue), $1.99$ (J-aggregate resonance, purple), and $1.94$\,eV (J-aggregate resonance, red).  Figure from Ref.\citenum{meza2021frenkel}}
    \label{fig:4}
\end{figure}

We have examined the theoretical concepts by means of two-dimensional coherent excitation spectroscopy on PBTTT, with structure depicted in Fig.~\ref{fig:1}(A), which we reported extensively in Ref.~\citenum{meza2021frenkel}. In that work, we identified spectral features associated with the 0--0 excitation peak of both the H- and J-aggregate components of the hybrid aggregate, with cross peaks reflecting spectral correlations due to their shared ground state. 
The origin of the H-aggregate vibronic progression was centered at 2.06\,eV, while a weaker peak at 1.99\,eV was assigned to the J-aggregate vibronic origin. By performing two-quantum coherence measurements, we found spectral signatures of both 2H and 2J biexcitons. A cross-section 
of the 2D spectral data along the two-quantum energy axis ($\hbar \omega_{2Q} - \hbar \omega_{diag} $) relative to the two-quantum diagonal ($\omega_{diag} \equiv \omega_{2Q} = 2 \omega_{1Q}$) is 
shown in Fig.~\ref{fig:4}.
We found that 2H biexcitons displayed \emph{attractive} biexciton binding with energy $-64 \pm 6$\,meV, where as 2J biexcitons displayed \emph{repulsive} correlations with binding energy $+106 \pm 6$\ meV; that is that the energy of the 2H-biexciton resonance is lower than twice the H-aggregate resonance energy, while the corresponding energy for the 2J-biexciton resonance is higher. 

We rationalized this observation as depicted in Fig.~\ref{fig:1}: two quantum interactions for excitons dispersed along the polymer backbone (J aggregates) are with $J < 0$, while $J > 0$ for those for between excitons dispersed on several chains (H aggregates). Considering physically reasonable parameters, we concluded that biexcitons in PBTTT are stable by the arguments depicted in Fig.~\ref{fig:4}. 

\section{Perspective}


We present here theoretical and experimental
evidence supporting the formation of bound
Frenkel biexcitons in a molecular aggregate
material.  In our theoretical analysis, we solved the full 2D interacting model and give the 
conditions necessary for the formation of stable, stationary states corresponding to bound exciton pairs.  The model provides a road-map for 
developing a bi-exciton materials genome in terms
of the properties of the J- and H-excitons.  
We show that for bound biexcitons, both the 
exciton-exciton interaction $U$ and the single-particle hopping integral $t$ must have the 
same sign. Furthermore,  $U/t > 2$ so that
the exciton/exciton potential interaction is greater than their total kinetic energy. 
Curiously, we find that while H-like excitons
form bound biexcitons with attractive interactions, J-like exciton pairs form bound states arising 
from a repulsive interaction.
Curiously, we find that while the 2H biexciton is stabilized by interactions with the 
lattice phonons, the 2J biexciton is destabilized to the extent that strong lattice/exciton 
interactions will produce only unbound biexciton states, which certainly complicates the 
observation of these states in systems with strong exciton/phonon coupling.  
Nonetheless, the 2J biexciton is clearly apparent in the 2D two-quantum experiments
 in Ref.\citenum{meza2021frenkel}.

An open question, however, is the nature of the
exciton-exciton interaction itself.  Here, we introduce it as a parameter into our model with a dipole-dipole like form in that the exciton-exciton coupling in one direction is different from that in a perpendicular direction.  It is also important to
point out that while this interaction appears to have a dipole-like form, it necessarily must reflect the {\em static} dipole of the local Frenkel excitons
rather than their transition dipole moments, which 
are responsible for the single exciton transfer between local sites.  Computing these interactions
from a first-principle {\em ab initio} theory remains a formidable challenge since it necessitates the accurate calculation of doubly excited states with some degree of charge-separation.

\begin{acknowledgments}

The work at the University of Houston was funded in
part by the  National Science Foundation (CHE-2102506) and the Robert A. Welch Foundation (E-1337). 
The work at Georgia Tech was funded by the National Science Foundation (DMR-1904293). %

\end{acknowledgments}

{\bf Author Declaration:}
Both authors contributed equally to the design and analysis of this
study.  Theoretical methodology was developed by ERB and experiments
were designed and analyzed by CS.  The manuscript was written 
and edited by ERB.
Both authors have read and agreed to the published version of the manuscript.
The authors have no conflicts to disclose.

{\bf Data Availability:}The data that supports the findings of this study are available within the article.

\bibliography{PBTTTLit,DavydovSoliton}

\begin{thebibliography}{30}%
\makeatletter
\providecommand \@ifxundefined [1]{%
 \@ifx{#1\undefined}
}%
\providecommand \@ifnum [1]{%
 \ifnum #1\expandafter \@firstoftwo
 \else \expandafter \@secondoftwo
 \fi
}%
\providecommand \@ifx [1]{%
 \ifx #1\expandafter \@firstoftwo
 \else \expandafter \@secondoftwo
 \fi
}%
\providecommand \natexlab [1]{#1}%
\providecommand \enquote  [1]{``#1''}%
\providecommand \bibnamefont  [1]{#1}%
\providecommand \bibfnamefont [1]{#1}%
\providecommand \citenamefont [1]{#1}%
\providecommand \href@noop [0]{\@secondoftwo}%
\providecommand \href [0]{\begingroup \@sanitize@url \@href}%
\providecommand \@href[1]{\@@startlink{#1}\@@href}%
\providecommand \@@href[1]{\endgroup#1\@@endlink}%
\providecommand \@sanitize@url [0]{\catcode `\\12\catcode `\$12\catcode
  `\&12\catcode `\#12\catcode `\^12\catcode `\_12\catcode `\%12\relax}%
\providecommand \@@startlink[1]{}%
\providecommand \@@endlink[0]{}%
\providecommand \url  [0]{\begingroup\@sanitize@url \@url }%
\providecommand \@url [1]{\endgroup\@href {#1}{\urlprefix }}%
\providecommand \urlprefix  [0]{URL }%
\providecommand \Eprint [0]{\href }%
\providecommand \doibase [0]{https://doi.org/}%
\providecommand \selectlanguage [0]{\@gobble}%
\providecommand \bibinfo  [0]{\@secondoftwo}%
\providecommand \bibfield  [0]{\@secondoftwo}%
\providecommand \translation [1]{[#1]}%
\providecommand \BibitemOpen [0]{}%
\providecommand \bibitemStop [0]{}%
\providecommand \bibitemNoStop [0]{.\EOS\space}%
\providecommand \EOS [0]{\spacefactor3000\relax}%
\providecommand \BibitemShut  [1]{\csname bibitem#1\endcsname}%
\let\auto@bib@innerbib\@empty
\bibitem [{\citenamefont {Frenkel}(1931)}]{frenkel1931transformation}%
  \BibitemOpen
  \bibfield  {author} {\bibinfo {author} {\bibfnamefont {J.}~\bibnamefont
  {Frenkel}},\ }\bibfield  {title} {\enquote {\bibinfo {title} {On the
  transformation of light into heat in solids. i},}\ }\href@noop {} {\bibfield
  {journal} {\bibinfo  {journal} {Phys. Rev.}\ }\textbf {\bibinfo {volume}
  {37}},\ \bibinfo {pages} {17} (\bibinfo {year} {1931})}\BibitemShut {NoStop}%
\bibitem [{\citenamefont {Agranovich}\ and\ \citenamefont
  {Toshich}(1968)}]{agranovich1968collective}%
  \BibitemOpen
  \bibfield  {author} {\bibinfo {author} {\bibfnamefont {V.}~\bibnamefont
  {Agranovich}}\ and\ \bibinfo {author} {\bibfnamefont {B.}~\bibnamefont
  {Toshich}},\ }\bibfield  {title} {\enquote {\bibinfo {title} {Collective
  properties of frenkel excitons},}\ }\href@noop {} {\bibfield  {journal}
  {\bibinfo  {journal} {Sov. Phys. JETP}\ }\textbf {\bibinfo {volume} {26}},\
  \bibinfo {pages} {104--112} (\bibinfo {year} {1968})}\BibitemShut {NoStop}%
\bibitem [{\citenamefont {Silva}\ \emph {et~al.}(2001)\citenamefont {Silva},
  \citenamefont {Dhoot}, \citenamefont {Russell}, \citenamefont {Stevens},
  \citenamefont {Arias}, \citenamefont {MacKenzie}, \citenamefont {Greenham},
  \citenamefont {Friend}, \citenamefont {Setayesh},\ and\ \citenamefont
  {M{\"u}llen}}]{silva2001efficient}%
  \BibitemOpen
  \bibfield  {author} {\bibinfo {author} {\bibfnamefont {C.}~\bibnamefont
  {Silva}}, \bibinfo {author} {\bibfnamefont {A.~S.}\ \bibnamefont {Dhoot}},
  \bibinfo {author} {\bibfnamefont {D.~M.}\ \bibnamefont {Russell}}, \bibinfo
  {author} {\bibfnamefont {M.~A.}\ \bibnamefont {Stevens}}, \bibinfo {author}
  {\bibfnamefont {A.~C.}\ \bibnamefont {Arias}}, \bibinfo {author}
  {\bibfnamefont {J.~D.}\ \bibnamefont {MacKenzie}}, \bibinfo {author}
  {\bibfnamefont {N.~C.}\ \bibnamefont {Greenham}}, \bibinfo {author}
  {\bibfnamefont {R.~H.}\ \bibnamefont {Friend}}, \bibinfo {author}
  {\bibfnamefont {S.}~\bibnamefont {Setayesh}},\ and\ \bibinfo {author}
  {\bibfnamefont {K.}~\bibnamefont {M{\"u}llen}},\ }\bibfield  {title}
  {\enquote {\bibinfo {title} {Efficient exciton dissociation via two-step
  photoexcitation in polymeric semiconductors},}\ }\href@noop {} {\bibfield
  {journal} {\bibinfo  {journal} {Phys. Rev. B}\ }\textbf {\bibinfo {volume}
  {64}},\ \bibinfo {pages} {125211} (\bibinfo {year} {2001})}\BibitemShut
  {NoStop}%
\bibitem [{\citenamefont {Stevens}\ \emph {et~al.}(2001)\citenamefont
  {Stevens}, \citenamefont {Silva}, \citenamefont {Russell},\ and\
  \citenamefont {Friend}}]{stevens2001exciton}%
  \BibitemOpen
  \bibfield  {author} {\bibinfo {author} {\bibfnamefont {M.~A.}\ \bibnamefont
  {Stevens}}, \bibinfo {author} {\bibfnamefont {C.}~\bibnamefont {Silva}},
  \bibinfo {author} {\bibfnamefont {D.~M.}\ \bibnamefont {Russell}},\ and\
  \bibinfo {author} {\bibfnamefont {R.~H.}\ \bibnamefont {Friend}},\ }\bibfield
   {title} {\enquote {\bibinfo {title} {Exciton dissociation mechanisms in the
  polymeric semiconductors poly (9, 9-dioctylfluorene) and poly (9,
  9-dioctylfluorene-co-benzothiadiazole)},}\ }\href@noop {} {\bibfield
  {journal} {\bibinfo  {journal} {Phys. Rev. B}\ }\textbf {\bibinfo {volume}
  {63}},\ \bibinfo {pages} {165213} (\bibinfo {year} {2001})}\BibitemShut
  {NoStop}%
\bibitem [{\citenamefont {Silva}\ \emph {et~al.}(2002)\citenamefont {Silva},
  \citenamefont {Russell}, \citenamefont {Dhoot}, \citenamefont {Herz},
  \citenamefont {Daniel}, \citenamefont {Greenham}, \citenamefont {Arias},
  \citenamefont {Setayesh}, \citenamefont {M{\"u}llen},\ and\ \citenamefont
  {Friend}}]{silva2002exciton}%
  \BibitemOpen
  \bibfield  {author} {\bibinfo {author} {\bibfnamefont {C.}~\bibnamefont
  {Silva}}, \bibinfo {author} {\bibfnamefont {D.~M.}\ \bibnamefont {Russell}},
  \bibinfo {author} {\bibfnamefont {A.~S.}\ \bibnamefont {Dhoot}}, \bibinfo
  {author} {\bibfnamefont {L.~M.}\ \bibnamefont {Herz}}, \bibinfo {author}
  {\bibfnamefont {C.}~\bibnamefont {Daniel}}, \bibinfo {author} {\bibfnamefont
  {N.~C.}\ \bibnamefont {Greenham}}, \bibinfo {author} {\bibfnamefont {A.~C.}\
  \bibnamefont {Arias}}, \bibinfo {author} {\bibfnamefont {S.}~\bibnamefont
  {Setayesh}}, \bibinfo {author} {\bibfnamefont {K.}~\bibnamefont
  {M{\"u}llen}},\ and\ \bibinfo {author} {\bibfnamefont {R.~H.}\ \bibnamefont
  {Friend}},\ }\bibfield  {title} {\enquote {\bibinfo {title} {Exciton and
  polaron dynamics in a step-ladder polymeric semiconductor: the influence of
  interchain order},}\ }\href@noop {} {\bibfield  {journal} {\bibinfo
  {journal} {J. Phys.: Condens. Matter}\ }\textbf {\bibinfo {volume} {14}},\
  \bibinfo {pages} {9803} (\bibinfo {year} {2002})}\BibitemShut {NoStop}%
\bibitem [{\citenamefont {Scholes}\ \emph {et~al.}(2011)\citenamefont
  {Scholes}, \citenamefont {Fleming}, \citenamefont {Olaya-Castro},\ and\
  \citenamefont {Van~Grondelle}}]{scholes2011lessons}%
  \BibitemOpen
  \bibfield  {author} {\bibinfo {author} {\bibfnamefont {G.~D.}\ \bibnamefont
  {Scholes}}, \bibinfo {author} {\bibfnamefont {G.~R.}\ \bibnamefont
  {Fleming}}, \bibinfo {author} {\bibfnamefont {A.}~\bibnamefont
  {Olaya-Castro}},\ and\ \bibinfo {author} {\bibfnamefont {R.}~\bibnamefont
  {Van~Grondelle}},\ }\bibfield  {title} {\enquote {\bibinfo {title} {Lessons
  from nature about solar light harvesting},}\ }\href@noop {} {\bibfield
  {journal} {\bibinfo  {journal} {Nat. Chem.}\ }\textbf {\bibinfo {volume}
  {3}},\ \bibinfo {pages} {763--774} (\bibinfo {year} {2011})}\BibitemShut
  {NoStop}%
\bibitem [{\citenamefont {Avanaki}\ and\ \citenamefont
  {Schatz}(2021)}]{n2021mechanistic}%
  \BibitemOpen
  \bibfield  {author} {\bibinfo {author} {\bibfnamefont {K.~N.}\ \bibnamefont
  {Avanaki}}\ and\ \bibinfo {author} {\bibfnamefont {G.~C.}\ \bibnamefont
  {Schatz}},\ }\bibfield  {title} {\enquote {\bibinfo {title} {Mechanistic
  understanding of entanglement and heralding in cascade emitters},}\
  }\href@noop {} {\bibfield  {journal} {\bibinfo  {journal} {J. Chem. Phys.}\
  }\textbf {\bibinfo {volume} {154}},\ \bibinfo {pages} {024304} (\bibinfo
  {year} {2021})}\BibitemShut {NoStop}%
\bibitem [{\citenamefont {Spano}, \citenamefont {Agranovich},\ and\
  \citenamefont {Mukamel}(1991)}]{spano1991biexciton}%
  \BibitemOpen
  \bibfield  {author} {\bibinfo {author} {\bibfnamefont {F.~C.}\ \bibnamefont
  {Spano}}, \bibinfo {author} {\bibfnamefont {V.}~\bibnamefont {Agranovich}},\
  and\ \bibinfo {author} {\bibfnamefont {S.}~\bibnamefont {Mukamel}},\
  }\bibfield  {title} {\enquote {\bibinfo {title} {Biexciton states and
  two-photon absorption in molecular monolayers},}\ }\href@noop {} {\bibfield
  {journal} {\bibinfo  {journal} {J. Chem. Phys.}\ }\textbf {\bibinfo {volume}
  {95}},\ \bibinfo {pages} {1400--1409} (\bibinfo {year} {1991})}\BibitemShut
  {NoStop}%
\bibitem [{\citenamefont {Vektaris}(1994)}]{Vektaris1994}%
  \BibitemOpen
  \bibfield  {author} {\bibinfo {author} {\bibfnamefont {G.}~\bibnamefont
  {Vektaris}},\ }\bibfield  {title} {\enquote {\bibinfo {title} {A new approach
  to the molecular biexciton theory},}\ }\href
  {https://doi.org/10.1063/1.467616} {\bibfield  {journal} {\bibinfo  {journal}
  {The Journal of Chemical Physics}\ }\textbf {\bibinfo {volume} {101}},\
  \bibinfo {pages} {3031--3040} (\bibinfo {year} {1994})},\ \Eprint
  {https://arxiv.org/abs/https://doi.org/10.1063/1.467616}
  {https://doi.org/10.1063/1.467616} \BibitemShut {NoStop}%
\bibitem [{\citenamefont {Guo}, \citenamefont {Chandross},\ and\ \citenamefont
  {Mazumdar}(1995)}]{guo1995stable}%
  \BibitemOpen
  \bibfield  {author} {\bibinfo {author} {\bibfnamefont {F.}~\bibnamefont
  {Guo}}, \bibinfo {author} {\bibfnamefont {M.}~\bibnamefont {Chandross}},\
  and\ \bibinfo {author} {\bibfnamefont {S.}~\bibnamefont {Mazumdar}},\
  }\bibfield  {title} {\enquote {\bibinfo {title} {Stable biexcitons in
  conjugated polymers},}\ }\href@noop {} {\bibfield  {journal} {\bibinfo
  {journal} {Phys. Rev. Lett.}\ }\textbf {\bibinfo {volume} {74}},\ \bibinfo
  {pages} {2086} (\bibinfo {year} {1995})}\BibitemShut {NoStop}%
\bibitem [{\citenamefont {Gallagher}\ and\ \citenamefont
  {Spano}(1996)}]{gallagher1996theory}%
  \BibitemOpen
  \bibfield  {author} {\bibinfo {author} {\bibfnamefont {F.~B.}\ \bibnamefont
  {Gallagher}}\ and\ \bibinfo {author} {\bibfnamefont {F.~C.}\ \bibnamefont
  {Spano}},\ }\bibfield  {title} {\enquote {\bibinfo {title} {Theory of
  biexcitons in one-dimensional polymers},}\ }\href@noop {} {\bibfield
  {journal} {\bibinfo  {journal} {Phys. Rev. B}\ }\textbf {\bibinfo {volume}
  {53}},\ \bibinfo {pages} {3790} (\bibinfo {year} {1996})}\BibitemShut
  {NoStop}%
\bibitem [{\citenamefont {Mazumdar}\ \emph {et~al.}(1996)\citenamefont
  {Mazumdar}, \citenamefont {Guo}, \citenamefont {Meissner}, \citenamefont
  {Fluegel},\ and\ \citenamefont {Peyghambarian}}]{mazumdar1996exciton}%
  \BibitemOpen
  \bibfield  {author} {\bibinfo {author} {\bibfnamefont {S.}~\bibnamefont
  {Mazumdar}}, \bibinfo {author} {\bibfnamefont {F.}~\bibnamefont {Guo}},
  \bibinfo {author} {\bibfnamefont {K.}~\bibnamefont {Meissner}}, \bibinfo
  {author} {\bibfnamefont {B.}~\bibnamefont {Fluegel}},\ and\ \bibinfo {author}
  {\bibfnamefont {N.}~\bibnamefont {Peyghambarian}},\ }\bibfield  {title}
  {\enquote {\bibinfo {title} {Exciton-to-biexciton transition in
  quasi-one-dimensional organics},}\ }\href@noop {} {\bibfield  {journal}
  {\bibinfo  {journal} {J. Chem. Phys.}\ }\textbf {\bibinfo {volume} {104}},\
  \bibinfo {pages} {9292--9296} (\bibinfo {year} {1996})}\BibitemShut {NoStop}%
\bibitem [{\citenamefont {Agranovich}\ \emph {et~al.}(2000)\citenamefont
  {Agranovich}, \citenamefont {Dubovsky}, \citenamefont {Basko}, \citenamefont
  {La~Rocca},\ and\ \citenamefont {Bassani}}]{agranovich2000kinematic}%
  \BibitemOpen
  \bibfield  {author} {\bibinfo {author} {\bibfnamefont {V.}~\bibnamefont
  {Agranovich}}, \bibinfo {author} {\bibfnamefont {O.}~\bibnamefont
  {Dubovsky}}, \bibinfo {author} {\bibfnamefont {D.}~\bibnamefont {Basko}},
  \bibinfo {author} {\bibfnamefont {G.}~\bibnamefont {La~Rocca}},\ and\
  \bibinfo {author} {\bibfnamefont {F.}~\bibnamefont {Bassani}},\ }\bibfield
  {title} {\enquote {\bibinfo {title} {Kinematic frenkel biexcitons},}\
  }\href@noop {} {\bibfield  {journal} {\bibinfo  {journal} {J. Lumin.}\
  }\textbf {\bibinfo {volume} {85}},\ \bibinfo {pages} {221--232} (\bibinfo
  {year} {2000})}\BibitemShut {NoStop}%
\bibitem [{\citenamefont {Kun}\ \emph {et~al.}(2009)\citenamefont {Kun},
  \citenamefont {Shi-Jie}, \citenamefont {Yuan}, \citenamefont {Sun},
  \citenamefont {De-Sheng},\ and\ \citenamefont {Xian}}]{kun2009effect}%
  \BibitemOpen
  \bibfield  {author} {\bibinfo {author} {\bibfnamefont {G.}~\bibnamefont
  {Kun}}, \bibinfo {author} {\bibfnamefont {X.}~\bibnamefont {Shi-Jie}},
  \bibinfo {author} {\bibfnamefont {L.}~\bibnamefont {Yuan}}, \bibinfo {author}
  {\bibfnamefont {Y.}~\bibnamefont {Sun}}, \bibinfo {author} {\bibfnamefont
  {L.}~\bibnamefont {De-Sheng}},\ and\ \bibinfo {author} {\bibfnamefont
  {Z.}~\bibnamefont {Xian}},\ }\bibfield  {title} {\enquote {\bibinfo {title}
  {Effect of interchain coupling on a biexciton in organic polymers},}\
  }\href@noop {} {\bibfield  {journal} {\bibinfo  {journal} {Chin. Phys. B}\
  }\textbf {\bibinfo {volume} {18}},\ \bibinfo {pages} {2961} (\bibinfo {year}
  {2009})}\BibitemShut {NoStop}%
\bibitem [{\citenamefont {Xiang}, \citenamefont {Litinskaya},\ and\
  \citenamefont {Krems}(2012)}]{xiang2012tunable}%
  \BibitemOpen
  \bibfield  {author} {\bibinfo {author} {\bibfnamefont {P.}~\bibnamefont
  {Xiang}}, \bibinfo {author} {\bibfnamefont {M.}~\bibnamefont {Litinskaya}},\
  and\ \bibinfo {author} {\bibfnamefont {R.~V.}\ \bibnamefont {Krems}},\
  }\bibfield  {title} {\enquote {\bibinfo {title} {Tunable exciton interactions
  in optical lattices with polar molecules},}\ }\href@noop {} {\bibfield
  {journal} {\bibinfo  {journal} {Phys. Rev. A}\ }\textbf {\bibinfo {volume}
  {85}},\ \bibinfo {pages} {061401} (\bibinfo {year} {2012})}\BibitemShut
  {NoStop}%
\bibitem [{\citenamefont {Chakrabarti}\ \emph {et~al.}(1998)\citenamefont
  {Chakrabarti}, \citenamefont {Schmidt}, \citenamefont {Valencia},
  \citenamefont {Fluegel}, \citenamefont {Mazumdar}, \citenamefont
  {Armstrong},\ and\ \citenamefont {Peyghambarian}}]{chakrabarti1998evidence}%
  \BibitemOpen
  \bibfield  {author} {\bibinfo {author} {\bibfnamefont {A.}~\bibnamefont
  {Chakrabarti}}, \bibinfo {author} {\bibfnamefont {A.}~\bibnamefont
  {Schmidt}}, \bibinfo {author} {\bibfnamefont {V.}~\bibnamefont {Valencia}},
  \bibinfo {author} {\bibfnamefont {B.}~\bibnamefont {Fluegel}}, \bibinfo
  {author} {\bibfnamefont {S.}~\bibnamefont {Mazumdar}}, \bibinfo {author}
  {\bibfnamefont {N.}~\bibnamefont {Armstrong}},\ and\ \bibinfo {author}
  {\bibfnamefont {N.}~\bibnamefont {Peyghambarian}},\ }\bibfield  {title}
  {\enquote {\bibinfo {title} {Evidence for exciton-exciton binding in a
  molecular aggregate},}\ }\href@noop {} {\bibfield  {journal} {\bibinfo
  {journal} {Phys. Rev. B}\ }\textbf {\bibinfo {volume} {57}},\ \bibinfo
  {pages} {R4206} (\bibinfo {year} {1998})}\BibitemShut {NoStop}%
\bibitem [{\citenamefont {Klimov}\ \emph {et~al.}(1998)\citenamefont {Klimov},
  \citenamefont {McBranch}, \citenamefont {Barashkov},\ and\ \citenamefont
  {Ferraris}}]{klimov1998biexcitons}%
  \BibitemOpen
  \bibfield  {author} {\bibinfo {author} {\bibfnamefont {V.~I.}\ \bibnamefont
  {Klimov}}, \bibinfo {author} {\bibfnamefont {D.}~\bibnamefont {McBranch}},
  \bibinfo {author} {\bibfnamefont {N.}~\bibnamefont {Barashkov}},\ and\
  \bibinfo {author} {\bibfnamefont {J.}~\bibnamefont {Ferraris}},\ }\bibfield
  {title} {\enquote {\bibinfo {title} {Biexcitons in $\pi$-conjugated
  oligomers: Intensity-dependent femtosecond transient-absorption study},}\
  }\href@noop {} {\bibfield  {journal} {\bibinfo  {journal} {Phys. Rev. B}\
  }\textbf {\bibinfo {volume} {58}},\ \bibinfo {pages} {7654} (\bibinfo {year}
  {1998})}\BibitemShut {NoStop}%
\bibitem [{\citenamefont {McCulloch}\ \emph {et~al.}(2006)\citenamefont
  {McCulloch}, \citenamefont {Heeney}, \citenamefont {Bailey}, \citenamefont
  {Genevicius}, \citenamefont {MacDonald}, \citenamefont {Shkunov},
  \citenamefont {Sparrowe}, \citenamefont {Tierney}, \citenamefont {Wagner},
  \citenamefont {Zhang} \emph {et~al.}}]{mcculloch2006liquid}%
  \BibitemOpen
  \bibfield  {author} {\bibinfo {author} {\bibfnamefont {I.}~\bibnamefont
  {McCulloch}}, \bibinfo {author} {\bibfnamefont {M.}~\bibnamefont {Heeney}},
  \bibinfo {author} {\bibfnamefont {C.}~\bibnamefont {Bailey}}, \bibinfo
  {author} {\bibfnamefont {K.}~\bibnamefont {Genevicius}}, \bibinfo {author}
  {\bibfnamefont {I.}~\bibnamefont {MacDonald}}, \bibinfo {author}
  {\bibfnamefont {M.}~\bibnamefont {Shkunov}}, \bibinfo {author} {\bibfnamefont
  {D.}~\bibnamefont {Sparrowe}}, \bibinfo {author} {\bibfnamefont
  {S.}~\bibnamefont {Tierney}}, \bibinfo {author} {\bibfnamefont
  {R.}~\bibnamefont {Wagner}}, \bibinfo {author} {\bibfnamefont
  {W.}~\bibnamefont {Zhang}}, \emph {et~al.},\ }\bibfield  {title} {\enquote
  {\bibinfo {title} {Liquid-crystalline semiconducting polymers with high
  charge-carrier mobility},}\ }\href@noop {} {\bibfield  {journal} {\bibinfo
  {journal} {Nat. Mater.}\ }\textbf {\bibinfo {volume} {5}},\ \bibinfo {pages}
  {328--333} (\bibinfo {year} {2006})}\BibitemShut {NoStop}%
\bibitem [{\citenamefont {Guti{\'e}rrez-Meza}\ \emph
  {et~al.}(2021)\citenamefont {Guti{\'e}rrez-Meza}, \citenamefont {Malatesta},
  \citenamefont {Li}, \citenamefont {Bargigia}, \citenamefont {Kandada},
  \citenamefont {Valverde-Ch{\'a}vez}, \citenamefont {Kim}, \citenamefont {Li},
  \citenamefont {Stingelin}, \citenamefont {Tretiak}, \citenamefont {Bittner},\
  and\ \citenamefont {Silva-Acu{\~n}a}}]{meza2021frenkel}%
  \BibitemOpen
  \bibfield  {author} {\bibinfo {author} {\bibfnamefont {E.}~\bibnamefont
  {Guti{\'e}rrez-Meza}}, \bibinfo {author} {\bibfnamefont {R.}~\bibnamefont
  {Malatesta}}, \bibinfo {author} {\bibfnamefont {H.}~\bibnamefont {Li}},
  \bibinfo {author} {\bibfnamefont {I.}~\bibnamefont {Bargigia}}, \bibinfo
  {author} {\bibfnamefont {A.~R.~S.}\ \bibnamefont {Kandada}}, \bibinfo
  {author} {\bibfnamefont {D.~A.}\ \bibnamefont {Valverde-Ch{\'a}vez}},
  \bibinfo {author} {\bibfnamefont {S.-M.}\ \bibnamefont {Kim}}, \bibinfo
  {author} {\bibfnamefont {H.}~\bibnamefont {Li}}, \bibinfo {author}
  {\bibfnamefont {N.}~\bibnamefont {Stingelin}}, \bibinfo {author}
  {\bibfnamefont {S.}~\bibnamefont {Tretiak}}, \bibinfo {author} {\bibfnamefont
  {E.~R.}\ \bibnamefont {Bittner}},\ and\ \bibinfo {author} {\bibfnamefont
  {C.}~\bibnamefont {Silva-Acu{\~n}a}},\ }\bibfield  {title} {\enquote
  {\bibinfo {title} {Frenkel biexcitons in hybrid hj photophysical
  aggregates},}\ }\href {https://doi.org/10.1126/sciadv.abi5197} {\bibfield
  {journal} {\bibinfo  {journal} {Science Advances}\ }\textbf {\bibinfo
  {volume} {7}},\ \bibinfo {pages} {eabi5197} (\bibinfo {year} {2021})},\
  \Eprint
  {https://arxiv.org/abs/https://www.science.org/doi/pdf/10.1126/sciadv.abi5197}
  {https://www.science.org/doi/pdf/10.1126/sciadv.abi5197} \BibitemShut
  {NoStop}%
\bibitem [{Note1()}]{Note1}%
  \BibitemOpen
  \bibinfo {note} {Note, that here $t = -\hbar ^2/2\mu $ carries units of
  $[{\protect \rm Energy}]\cdot [{\protect \rm length}]^2$ and $U$ carries
  units of $[{\protect \rm Energy}]\cdot [{\protect \rm length}]^{-1}$ so that
  $\lambda $ carries the appropriate units of [length].}\BibitemShut {Stop}%
\bibitem [{\citenamefont {Scott}(1992)}]{SCOTT19921}%
  \BibitemOpen
  \bibfield  {author} {\bibinfo {author} {\bibfnamefont {A.}~\bibnamefont
  {Scott}},\ }\bibfield  {title} {\enquote {\bibinfo {title} {Davydov's
  soliton},}\ }\href
  {https://doi.org/https://doi.org/10.1016/0370-1573(92)90093-F} {\bibfield
  {journal} {\bibinfo  {journal} {Physics Reports}\ }\textbf {\bibinfo {volume}
  {217}},\ \bibinfo {pages} {1--67} (\bibinfo {year} {1992})}\BibitemShut
  {NoStop}%
\bibitem [{\citenamefont {Edler}\ \emph {et~al.}(2004)\citenamefont {Edler},
  \citenamefont {Pfister}, \citenamefont {Pouthier}, \citenamefont {Falvo},\
  and\ \citenamefont {Hamm}}]{Edler2004DirectOO}%
  \BibitemOpen
  \bibfield  {author} {\bibinfo {author} {\bibfnamefont {J.}~\bibnamefont
  {Edler}}, \bibinfo {author} {\bibfnamefont {R.}~\bibnamefont {Pfister}},
  \bibinfo {author} {\bibfnamefont {V.}~\bibnamefont {Pouthier}}, \bibinfo
  {author} {\bibfnamefont {C.}~\bibnamefont {Falvo}},\ and\ \bibinfo {author}
  {\bibfnamefont {P.}~\bibnamefont {Hamm}},\ }\bibfield  {title} {\enquote
  {\bibinfo {title} {Direct observation of self-trapped vibrational states in
  $\alpha$-helices},}\ }\href@noop {} {\bibfield  {journal} {\bibinfo
  {journal} {Physical Review Letters}\ }\textbf {\bibinfo {volume} {93}},\
  \bibinfo {pages} {106405} (\bibinfo {year} {2004})}\BibitemShut {NoStop}%
\bibitem [{\citenamefont {Sun}, \citenamefont {Luo},\ and\ \citenamefont
  {Zhao}(2010)}]{PhysRevB.82.014305}%
  \BibitemOpen
  \bibfield  {author} {\bibinfo {author} {\bibfnamefont {J.}~\bibnamefont
  {Sun}}, \bibinfo {author} {\bibfnamefont {B.}~\bibnamefont {Luo}},\ and\
  \bibinfo {author} {\bibfnamefont {Y.}~\bibnamefont {Zhao}},\ }\bibfield
  {title} {\enquote {\bibinfo {title} {Dynamics of a one-dimensional holstein
  polaron with the davydov ans\"atze},}\ }\href
  {https://doi.org/10.1103/PhysRevB.82.014305} {\bibfield  {journal} {\bibinfo
  {journal} {Phys. Rev. B}\ }\textbf {\bibinfo {volume} {82}},\ \bibinfo
  {pages} {014305} (\bibinfo {year} {2010})}\BibitemShut {NoStop}%
\bibitem [{\citenamefont {Goj}\ and\ \citenamefont
  {Bittner}(2011)}]{doi:10.1063/1.3592155}%
  \BibitemOpen
  \bibfield  {author} {\bibinfo {author} {\bibfnamefont {A.}~\bibnamefont
  {Goj}}\ and\ \bibinfo {author} {\bibfnamefont {E.~R.}\ \bibnamefont
  {Bittner}},\ }\bibfield  {title} {\enquote {\bibinfo {title} {Mixed quantum
  classical simulations of excitons in peptide helices},}\ }\href
  {https://doi.org/10.1063/1.3592155} {\bibfield  {journal} {\bibinfo
  {journal} {The Journal of Chemical Physics}\ }\textbf {\bibinfo {volume}
  {134}},\ \bibinfo {pages} {205103} (\bibinfo {year} {2011})},\ \Eprint
  {https://arxiv.org/abs/https://doi.org/10.1063/1.3592155}
  {https://doi.org/10.1063/1.3592155} \BibitemShut {NoStop}%
\bibitem [{\citenamefont {Goj}\ and\ \citenamefont {Bittner}(2010)}]{Goj:10}%
  \BibitemOpen
  \bibfield  {author} {\bibinfo {author} {\bibfnamefont {A.}~\bibnamefont
  {Goj}}\ and\ \bibinfo {author} {\bibfnamefont {E.~R.}\ \bibnamefont
  {Bittner}},\ }\bibfield  {title} {\enquote {\bibinfo {title} {Mixed quantum
  classical simulations of vibrational excitations in peptide helices},}\ }in\
  \href {https://doi.org/10.1364/LS.2010.LTuC1} {\emph {\bibinfo {booktitle}
  {Frontiers in Optics 2010/Laser Science XXVI}}}\ (\bibinfo  {publisher}
  {Optical Society of America},\ \bibinfo {year} {2010})\ p.\ \bibinfo {pages}
  {LTuC1}\BibitemShut {NoStop}%
\bibitem [{\citenamefont {Georgiev}\ and\ \citenamefont
  {Glazebrook}(2019{\natexlab{a}})}]{GEORGIEV2019257}%
  \BibitemOpen
  \bibfield  {author} {\bibinfo {author} {\bibfnamefont {D.~D.}\ \bibnamefont
  {Georgiev}}\ and\ \bibinfo {author} {\bibfnamefont {J.~F.}\ \bibnamefont
  {Glazebrook}},\ }\bibfield  {title} {\enquote {\bibinfo {title} {On the
  quantum dynamics of davydov solitons in protein $\alpha$-helices},}\ }\href
  {https://doi.org/https://doi.org/10.1016/j.physa.2018.11.026} {\bibfield
  {journal} {\bibinfo  {journal} {Physica A: Statistical Mechanics and its
  Applications}\ }\textbf {\bibinfo {volume} {517}},\ \bibinfo {pages}
  {257--269} (\bibinfo {year} {2019}{\natexlab{a}})}\BibitemShut {NoStop}%
\bibitem [{\citenamefont {Georgiev}\ and\ \citenamefont
  {Glazebrook}(2019{\natexlab{b}})}]{GEORGIEV2019275}%
  \BibitemOpen
  \bibfield  {author} {\bibinfo {author} {\bibfnamefont {D.~D.}\ \bibnamefont
  {Georgiev}}\ and\ \bibinfo {author} {\bibfnamefont {J.~F.}\ \bibnamefont
  {Glazebrook}},\ }\bibfield  {title} {\enquote {\bibinfo {title} {Quantum
  tunneling of davydov solitons through massive barriers},}\ }\href
  {https://doi.org/https://doi.org/10.1016/j.chaos.2019.04.013} {\bibfield
  {journal} {\bibinfo  {journal} {Chaos, Solitons and Fractals}\ }\textbf
  {\bibinfo {volume} {123}},\ \bibinfo {pages} {275--293} (\bibinfo {year}
  {2019}{\natexlab{b}})}\BibitemShut {NoStop}%
\bibitem [{\citenamefont {Kerr}\ and\ \citenamefont
  {Lomdahl}(1987)}]{PhysRevB.35.3629}%
  \BibitemOpen
  \bibfield  {author} {\bibinfo {author} {\bibfnamefont {W.~C.}\ \bibnamefont
  {Kerr}}\ and\ \bibinfo {author} {\bibfnamefont {P.~S.}\ \bibnamefont
  {Lomdahl}},\ }\bibfield  {title} {\enquote {\bibinfo {title}
  {Quantum-mechanical derivation of the equations of motion for davydov
  solitons},}\ }\href {https://doi.org/10.1103/PhysRevB.35.3629} {\bibfield
  {journal} {\bibinfo  {journal} {Phys. Rev. B}\ }\textbf {\bibinfo {volume}
  {35}},\ \bibinfo {pages} {3629--3632} (\bibinfo {year} {1987})}\BibitemShut
  {NoStop}%
\bibitem [{\citenamefont {Davydov}(1977)}]{DAVYDOV1977379}%
  \BibitemOpen
  \bibfield  {author} {\bibinfo {author} {\bibfnamefont {A.}~\bibnamefont
  {Davydov}},\ }\bibfield  {title} {\enquote {\bibinfo {title} {Solitons and
  energy transfer along protein molecules},}\ }\href
  {https://doi.org/https://doi.org/10.1016/0022-5193(77)90178-3} {\bibfield
  {journal} {\bibinfo  {journal} {Journal of Theoretical Biology}\ }\textbf
  {\bibinfo {volume} {66}},\ \bibinfo {pages} {379--387} (\bibinfo {year}
  {1977})}\BibitemShut {NoStop}%
\bibitem [{\citenamefont {Davydov}(1973)}]{DAVYDOV1973559}%
  \BibitemOpen
  \bibfield  {author} {\bibinfo {author} {\bibfnamefont {A.}~\bibnamefont
  {Davydov}},\ }\bibfield  {title} {\enquote {\bibinfo {title} {The theory of
  contraction of proteins under their excitation},}\ }\href
  {https://doi.org/https://doi.org/10.1016/0022-5193(73)90256-7} {\bibfield
  {journal} {\bibinfo  {journal} {Journal of Theoretical Biology}\ }\textbf
  {\bibinfo {volume} {38}},\ \bibinfo {pages} {559--569} (\bibinfo {year}
  {1973})}\BibitemShut {NoStop}%
\end{thebibliography}%

\end{document}